\documentclass[letterpaper, 10 pt, conference]{ieeeconf}  

\usepackage{color}
\usepackage{multirow}
\usepackage{booktabs}
\usepackage{array}
\usepackage{graphicx}
\usepackage{amsmath}
\usepackage{booktabs}  
\usepackage{array}
\usepackage{cite}
\usepackage{subfig}
\usepackage{url}
\usepackage{float}
\usepackage{hyperref}
\usepackage[table]{xcolor}
\newcolumntype{C}[1]{>{\centering\arraybackslash}m{#1}}

\IEEEoverridecommandlockouts                              

\title{\LARGE \bf
MealMeter: Using Multimodal Sensing and Machine Learning for Automatically Estimating Nutrition Intake
}

\author{Asiful Arefeen$^{1,2}$, Samantha Fessler$^{1}$, Sayyed Mostafa Mostafavi$^{1}$, Carol S Johnston$^{1}$ and Hassan Ghasemzadeh$^{1}$
\thanks{$^{1}$College of Health Solutions, Arizona State University, Phoenix, AZ 85004, USA}%
\thanks{$^{2}$School of Computing and Augmented Intelligence, Arizona State University, Tempe, AZ 85281, USA}%
\thanks{{email: \textcolor{blue}{aarefeen@asu.edu}}}
\thanks{{code available at: \textcolor{blue}{\href{https://github.com/Arefeen06088/MealMeter}{https://github.com/Arefeen06088/MealMeter}}}}
}

\begin{document}

\maketitle
\thispagestyle{empty}
\pagestyle{empty}

\begin{abstract}

Accurate estimation of meal macronutrient composition is a pre-perquisite for precision nutrition, metabolic health monitoring, and glycemic management. Traditional dietary assessment methods, such as self-reported food logs or diet recalls are time-intensive and prone to inaccuracies and biases. Several existing AI-driven frameworks are data intensive. In this study, we propose MealMeter, a machine learning driven method that leverages multimodal sensor data of wearable and mobile devices. Data are collected from $12$ participants to estimate macronutrient intake. Our approach integrates physiological signals (e.g., continuous glucose, heart rate variability), inertial motion data, and environmental cues to model the relationship between meal intake and metabolic responses. Using lightweight machine learning models trained on a diverse dataset of labeled meal events, MealMeter predicts the composition of carbohydrates, proteins, and fats with high accuracy. Our results demonstrate that multimodal sensing combined with machine learning significantly improves meal macronutrient estimation compared to the baselines including foundation model and achieves average mean absolute errors (MAE) and average root mean squared relative errors (RMSRE) as low as $13.2$ grams and $0.37$, respectively, for carbohydrates. Therefore, our developed system has the potential to automate meal tracking, enhance dietary interventions, and support personalized nutrition strategies for individuals managing metabolic disorders such as diabetes and obesity.
\newline

\indent \textit{Clinical relevance}— MealMeter is a clinically relevant tool for precision nutrition, glycemic management and obesity management because it offers accurate meal tracking through multimodal sensing and machine learning.
\end{abstract}

\section{Introduction}

\begin{figure*}[!tbh]
\centering
\includegraphics[width=\linewidth]{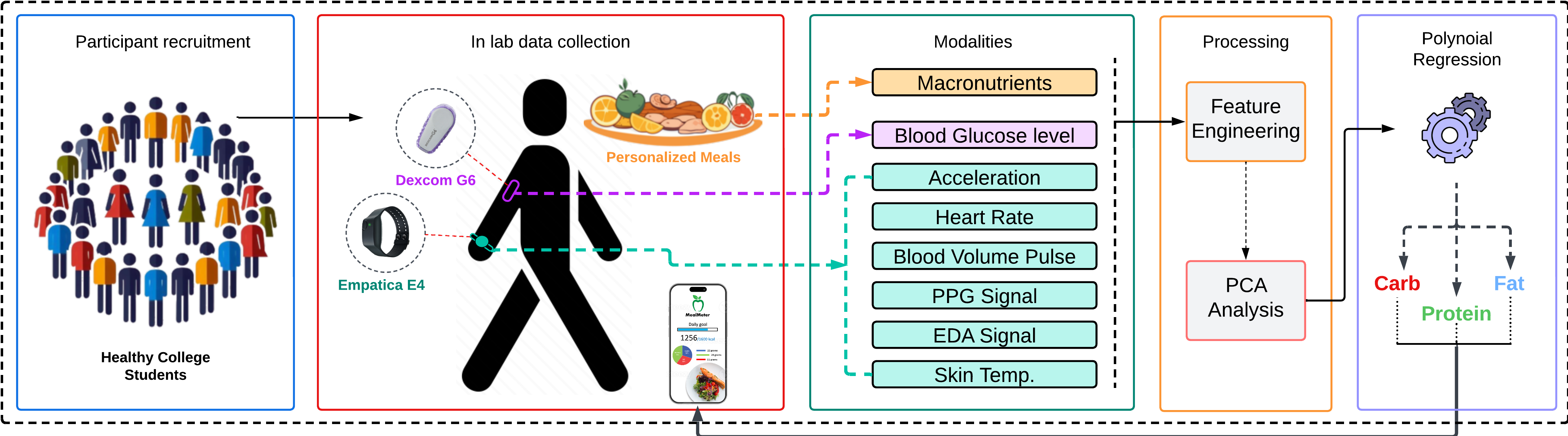}
\caption{Pipeline: MealMeter performs multi-modal fusion using data from Dexcom G6 CGM and Empatica E4 followed by feature engineering and light-weight model training for dietary assessment.}
\label{glytwin_dia}
\end{figure*}

Managing obesity, diabetes and cardiovascular disease (CVD) remains a continuous challenge for Americans. The age-adjusted prevalence of obesity among U.S. adults was 41.8\%, and severe obesity registered at 9.2 according to prepandemic national survey data collected from 2017 to 2020 \cite{Bryan2021NationalHA}. Obesity is closely associated with other serious chronic health conditions; for instance, almost 23\% of U.S. adults with obesity have diabetes while 58\% have high blood pressure, which is a major risk factor for CVDs \cite{Bryan2021NationalHA}. Addressing the root cause—obesity—requires a proactive approach to controlling consumption patterns and adopting healthy lifestyles \cite{Arefeen2022ComputationalFF}.

Meal monitoring facilitates the management and prevention the aforementioned chronic diseases. Accurate tracking of dietary intake enables individuals and healthcare providers  to make informed decisions about self-nutrition, which is vital for weight management, blood sugar control and reduction of CVD risk factors. Traditional dietary assessment techniques like self-reported food diaries are prone to inaccuracies due to inaccurate memory retention and subjective reporting \cite{Lo2024AIenabledWC}. However, advanced digital health technologies, including wearable devices and artificial intelligence (AI), offer innovative solutions for more precise and real-time dietary monitoring and to reduce laborious food logging \cite{Zheng2023ArtificialIA}.

Existing wearable technologies record physiological signals including heart rate variability, skin temperature and physical activity levels. Although these metrics offer indirect insights into metabolic responses to food intake, exact tracking of protein and fat consumption has remained challenging to date. Recent research explored the feasibility of advanced sensors and AI to estimate macronutrient intake by analyzing patterns in physiological data \cite{Zheng2023ArtificialIA}. Continuous glucose monitors (CGMs), for instance, can provide insights into blood sugar spikes due to carbohydrate intakes \cite{Cappon2019ContinuousGM}. High intakes of fat and high glycemic carbohydrates reduce heart rate variability \cite{Young2018HeartrateVA}. Prior research \cite{Moslehi2024QualityAQ} aimed to develop sophisticated methods to assess the impact of various macronutrients on physiological parameters, which could eventually lead to more precise dietary monitoring tools through wearable sensing.

Recent research focused on leveraging AI and wearable devices to enhance dietary assessment accuracy, predict abnormal health events and provide interventions \cite{Arefeen2022ForewarningPH, Arefeen2023GlySimMA, Arefeen2023DesigningUB, Shroff2023GlucoseAssistPB}. Zhang et al. \cite{Zhang2023JointEO} created embedding from food photographs using vision transformers, fused them with corresponding glycemic response embeddings and fed these patches to a regression model for calorie counting. Shao et al. \cite{Shao2023VisionbasedFN} used Red, Green, Blue-Depth of food images for precise estimation of portion sizes. Speech2Health \cite{Hezarjaribi2018Speech2HealthAM} used Natural Language Processing (NLP) techniques to convert spoken language to text and search the user-reported food name in a reference nutrition database to compute calorie intake. In another study, a five-point Gaussian filter was applied and area under the curve (AUC) was computed to capture the trend of the post-meal glycemic response \cite{Huo2019PredictingTM}. Later, this trend was fed to a multi-task neural network for carbohydrate, protein and fat estimation. Several techniques have been also proposed to accurately detect eating events leveraging sensor systems, although these techniques could not provide solutions for meal tracking, \cite{Heremans2020ArtificialNN, Farooq2019ValidationOS, Doulah2020AutomaticIM}.

MealMeter is the first of its kind to integrate CGM and wristband-based multimodal sensing (Figure~\ref{glytwin_dia}) to analyze the relationship between meal macronutrients and physiological signals for dietary assessment. Leveraging this synergy with computationally simple machine learning, MealMeter offers a fast, comprehensive and data-driven approach to tracking dietary intake and metabolic response. Therefore, contributions made by MealMeter can be summarized as follows-

\begin{itemize}
    \item For the purpose of developing MealMeter, we collected data from $12$ healthy individuals equipped with CGM sensor and Empatica E4 to capture metabolic response to customized meals for three non-consecutive days.
    \item We used the data from the user study to train lightweight regression models to map meal macronutrients from physiological signals and explore associations among these parameters.
\end{itemize}

\section{Methods}

\subsection{Data Collection}

\begin{figure*}[t]
\centering
\includegraphics[width=0.8\linewidth]{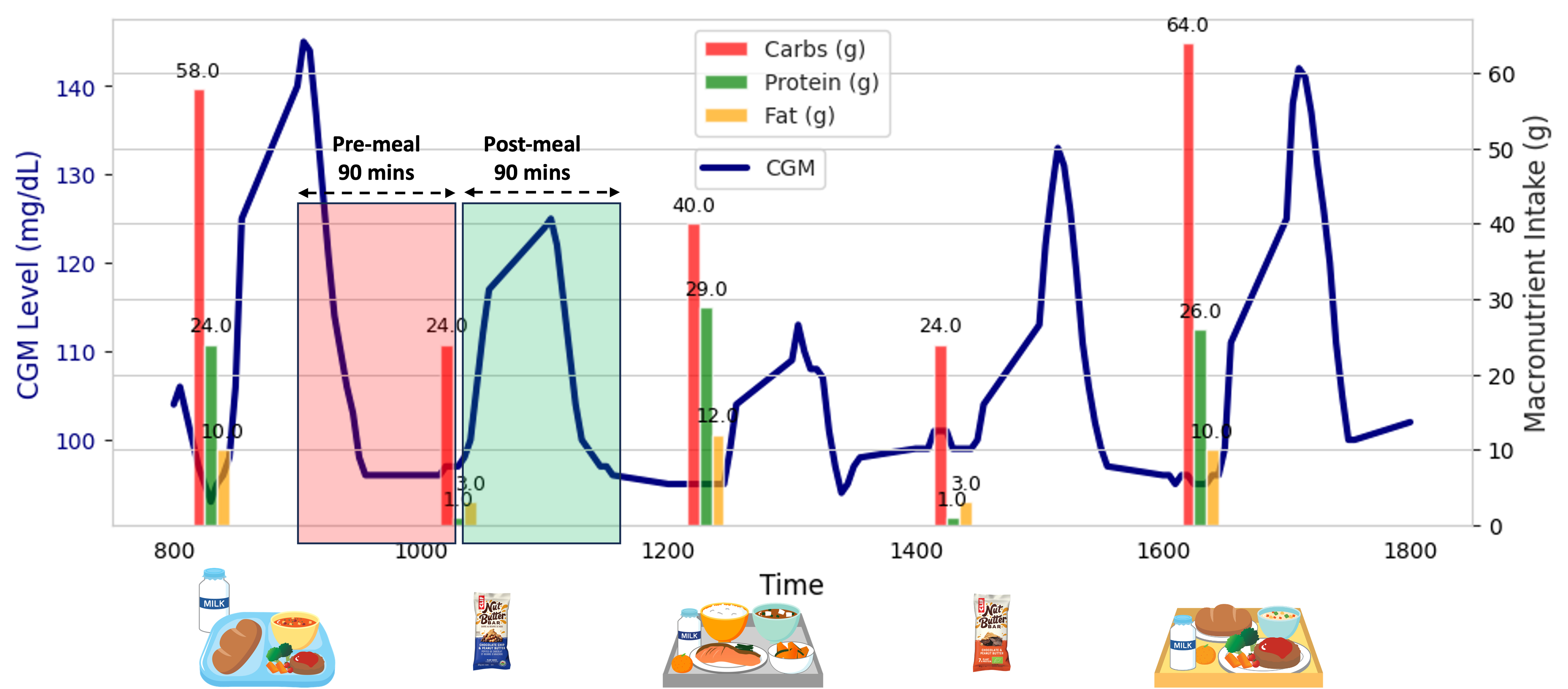}
\caption{Plot shows collected CGM signal from a subject as a response to the food items fed at different timestamps. It also shows 90-minute long pre- and post-meal windows as inputs to the pipeline.}
\label{data_dia}
\end{figure*}

Data were collected from an ongoing trial where we recruited $12$ healthy adults aged $23-31$ years with BMI $20.08-31.84$ to record device-derived responses to standardized meals. Participants had no diagnosed chronic metabolic or thyroid conditions, did not use recreational drugs, limited alcohol consumption to no more than two servings per day, and were not involved in competitive athletics or resistance training. All participants provided written consent to the study. The trial was approved by the Institutional Review Board at Arizona State University (IRB \#15102).

This study had three non-consecutive laboratory monitoring days. Upon enrollment, participants were equipped with wearable devices, including the Dexcom G6 CGM and the Empatica E4 wristband worn on their dominant arm. The CGM device collected blood glucose concentrations (BGL) at 5-minute intervals while the E4 recorded 3-axis acceleration (Acc), electrodermal activity (EDA), heart rate (HR), blood volume pulse (BVP) and skin temperature (TEMP) at different sampling frequencies ranging from 4Hz to 64Hz. Trained research personnel provided instructions on device usage and study study-related dos and don’ts.

During the three 10-hour laboratory monitoring sessions (8:00 AM–6:00 PM), participants arrived after a 12-hour overnight fast and received customized meals tailored to their resting energy requirements, calculated using the Mifflin-St Jeor equation \cite{Mifflin1990ANP}. Meals were classified as hypercaloric, eucaloric, or hypocaloric and consisted of commercial foods and frozen entre´es with consistent macronutrient distributions aligned with the acceptable macronutrient distribution range recommendations for adults aged 18–40 years: approximately $20\%$ protein, $55\%$ carbohydrate, and $25\%$ fat.

Meals were provided at 4-hour intervals (8:30 AM, 12:30 PM, and 4:30 PM) with two additional snacks at 10:30 AM and at 2:30 PM. A study team member present during laboratory visits recorded the precise meal times. Participants also responded to prompts on their smartphones at 30-minute intervals to record activity behavior which ensured sedentary condition and little effect on the glycemic response. 

Figure~\ref{data_dia} shows the meal events and the resulting glycemic response of a subject from a random study day.

\subsection{Feature Extraction}
MealMeter presents a systematic pipeline that processes the collected multimodal physiological data and maps them to meal macronutrient contents using a regression model. The pipeline includes data preprocessing, feature extraction in both time and frequency domains and dimensionality reduction via Principal Component Analysis (PCA).

A 90-minute input horizon is defined, meaning that all physiological signals recorded within a 90-minute window following meal intake are used for feature extraction and modeling. In addition to the post-meal signals, we add pre-meal blood glucose concentration. This segment of signal allows the model to account for carry-over effects from prior meals, as pre-meal glycemic status can influence postprandial responses.

To ensure uniformity in the dataset, all signals are re-sampled to a frequency of 8 Hz. A moving average filter with a window size of 20 samples is applied on acceleration magnitude to reduce noise and smoothen high-frequency variations while preserving underlying trends. A min-based normalization is applied to the blood glucose signals, where the signal is adjusted relative to the minimum observed level to improve comparability across subjects.

Next, we extract time-domain and frequency-domain features from the signals and translate them into meaningful numerical representations for machine learning models.

\begin{table}[h]
    \centering
    \caption{Extracted time domain and frequency domain features from the signals}
    {\renewcommand{\arraystretch}{1.2}
    \begin{tabular}{|C{5.2cm}|C{2.5cm}|}
        \hline
        \textbf{Time-Domain} & \textbf{Frequency-Domain} \\ 
        \hline
        \hline
        Min, Max, Skewness, Mean, SD, Kurtosis, Range, Root Mean Square (RMS), Median, Autocorrelation, Interquartile Range (IQR), Entropy, Zero-Crossing Rate (ZCR) & Power Spectral Density, Dominant Frequency, Spectral Entropy \\
        \hline
    \end{tabular}}
    \label{tab:features}
\end{table}

Once feature engineering is done, all features are standardized using $z$-score normalization:
\begin{align*}
X_{scaled} = \frac{X-\mu}{\sigma}
\end{align*}
where $\mu$ and $\sigma$ are the mean and standard deviation across the dataset. To reduce the dimension of the feature space without compromising information,  Principal Component Analysis (PCA) is applied with three principal components
\begin{align*}
    Z = X_{scaled}W
\end{align*}
where $W$ is the transformation matrix learned from PCA.
\subsection{Model Development}
Finally, the PCA-transformed features are used to train a linear regression model for meal macronutrient estimation. The linear regression model can be characterized as-
\begin{align}
    y = \beta_0 + \sum_i \beta_iz_k+\epsilon
\end{align}
where y represents macronutrient values (e.g., carbohydrate, fat, protein), 
$z_k$ are the PCA transformed features, $\beta_k$ are the regression coefficients, and $\epsilon$ is the error term.

All experiments are conducted using an 80/20 train-test split.

\subsection{Signal contributions}
Since we extract 16 statistical features from each physiological signal \( S_j \), contributions are initially estimated for each feature by multiplying principal component loadings \cite{Nakanishi2024PCaLDIES} and regression coefficients. Then contributions of features belonging to each signal are summed to compute the individual signal contribution-
\begin{align}
\gamma_i  & = \sum_{k=1}^{K} w_{ik} \beta_k \\ 
\Gamma_j  & = \sum_{i \in S_j}\gamma_i  = \sum_{i \in S_j} \sum_{k=1}^{K} w_{ik} \beta_k
\end{align}

where, \( w_{ik} \) is the weight loading of \( i^{th} \) feature in the \( k^{th} \) principal component, \( \gamma_i \) quantifies the importance of the \( i^{th} \) feature, \( \Gamma_j \) is the overall contribution of signal \( S_j \).

\subsection{Evaluation \& Baselines}

We evaluate and compare the results with the mean absolute error (MAE),
root mean square relative error (RMSRE) as well as the Pearson correlation (\textit{r}). MAE and RMSRE can be calculated as follows-
\begin{align*}
    \text{MAE}&=\frac{1}{n} \sum_{i=1}^{n} |y_i-\hat{y_i}| \;\;\;\;
    \text{RMSRE}&= \sqrt{\frac{1}{n} \sum_{i=1}^{n}{ \frac{ (y_i-\hat{y_i})^2 }{{y_i}^2} }}
\end{align*}
where $n$ is the number of samples, $y$ the true quantity and $\hat{y}$ is the estimate. 

We have found the following meal macronutrient estimation methods to compare against-

\textbf{\textit{Huo et al.}} \cite{Huo2019PredictingTM} applied Gaussian kernels at five equidistant time points to compute the AUC and capture the trend of the glycemic response before feeding them to a multi-task learning neural network.

\textbf{\textit{Yang et al.}} \cite{Yang2021AML} trained a Siamese neural network (SNN) to map glycemic responses into an embedding space. The model was optimized using a contrastive loss function that minimizes the difference between meal pairs in both the embedding space and the true macronutrient space.

\textit{\textbf{TabPFN}} \cite{hollmann2023tabpfn} is a transformer based foundation model with fast inference, specifically designed for tabular classification and regression. We fine tuned TabPFN on the PCA transformed data.

\section{Results}

We conducted experiments using data collected from $12$ participants to estimate carbohydrate, protein, and fat intake. First, subject-specific models have been developed and evaluated to capture individual variations in metabolic responses and dietary intake patterns. Table~\ref{individual_results} summarizes the results of subject-specific models.

For carbohydrate intake, MAE values ranged from $3.96$ g to $32.38$ g, with an average of $17.64$ g. The RMSRE remained relatively low with an average of $0.37$, which indicates stable performance. 

Protein estimation showed a wider spread in errors with MAE values between $2.93$ g and $18.4$ g and averaged $12.23$ g. However, RMSRE values varied significantly, with an average of $3.82$ which suggests that protein intake estimation using multi-modal sensor system is more challenging compared to estimating carbohydrate.
  
MAE for fat estimation was between $1.03$ g and $11.35$ g, with an average of $4.75$ g. The RMSRE values displayed considerable variation across subjects, ranging from $0.09$ to $4.6$, with an average of $0.86$. Despite this variability, the overall results indicate that the models performed relatively well in estimating carbohydrate and fat intake, while protein estimation appears to be more complex and very specific to individual subjects.

\begin{figure}[b!]
\centering
\includegraphics[width=\linewidth]{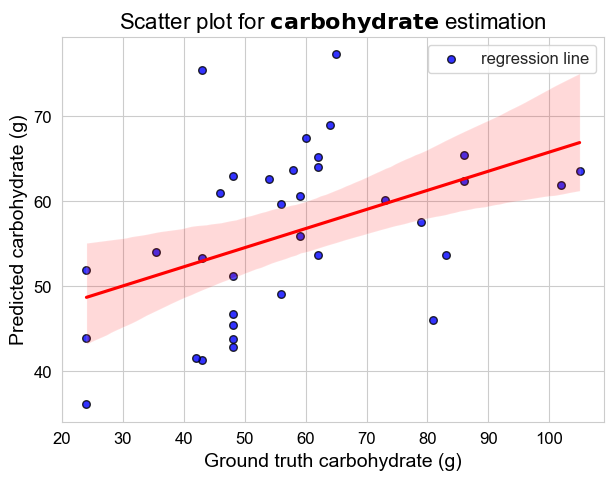}
\caption{Estimated carbohydrate vs. actual carbohydrate where correlation is $0.42$.}
\label{r_plot}
\end{figure}

\begin{figure*}[t!]
\centering
\subfloat{{\includegraphics[scale=0.32]{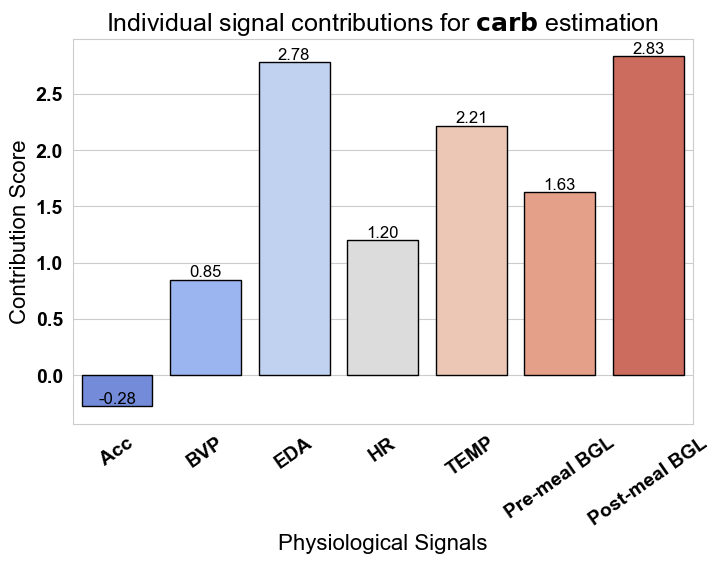}}}%
\subfloat{{\includegraphics[scale=0.32]{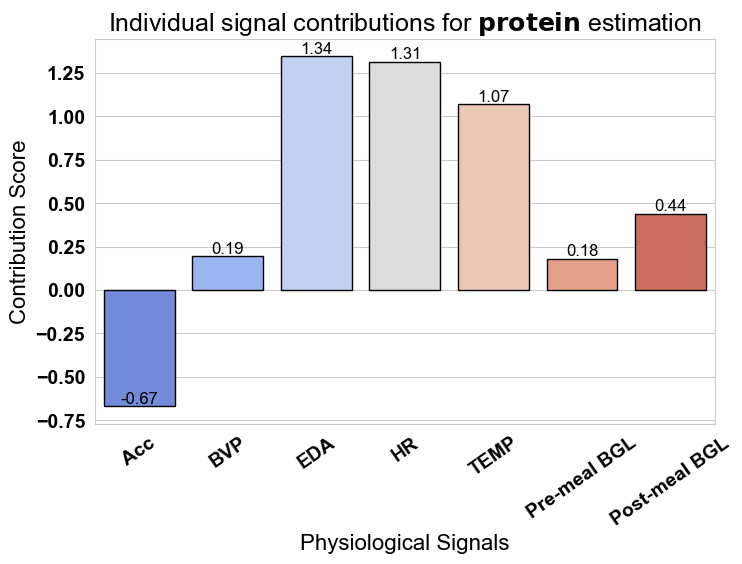}}}%
\subfloat{{\includegraphics[scale=0.32]{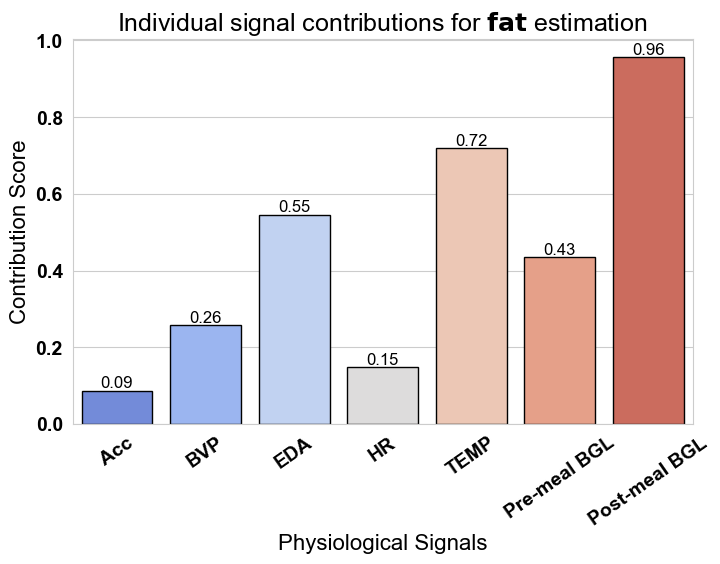}}}%
\caption{Contributions of individual physiological signals for Carbohydrate, Protein and Fat estimation.}
\label{contribution}%
\end{figure*}

\begin{table}[!h]
\centering
\footnotesize
\caption{MAE (grams) and RMSRE of estimated carbohydrate, protein and fat amounts for different subjects.}
{\renewcommand{\arraystretch}{1.2}
\begin{tabular}{|C{0.8cm}|C{0.7cm}C{0.9cm}|C{0.7cm}C{0.9cm}|C{0.7cm}C{0.9cm}|}
\toprule
   & \multicolumn{2}{c|}{\textbf{Carbohydrate}} & \multicolumn{2}{c|}{\textbf{Protein}} & \multicolumn{2}{c|}{\textbf{Fat}} \\ \hline
  \textbf{Subject} & \cellcolor{blue!18}\textbf{MAE} & \cellcolor{gray!18}\textbf{RMSRE} & \cellcolor{blue!18}\textbf{MAE} & \cellcolor{gray!18}\textbf{RMSRE} & \cellcolor{blue!18}\textbf{MAE} & \cellcolor{gray!18}\textbf{RMSRE} \\ \hline
P1 & 24.12 & 0.28 & 7.55 & 0.37 & 1.03 & 0.09 \\ \hline
P2 & 18.57 & 0.88 & 11.77 & 5.96 & 3.36 & 1.03 \\ \hline
P3 & 10.54 & 0.2 & 18.45 & 8.54 & 2.77 & 0.53 \\\hline
P4 & 27.34 & 0.68 & 13.59 & 7.37 & 6.87 & 1.08 \\\hline
P5 & 3.96 & 0.08 & 7.51 & 0.37 & 3.59 & 0.25 \\ \hline
P6 & 6.58 & 0.17 & 12.16 & 4.2 & 2.87 & 0.32\\\hline
P7 & 19.21 & 0.26 & 14.08 & 0.44 & 5.48 & 0.32\\\hline
P8 & 10.92 & 0.19 & 14.5 & 6.36 & 5.12 & 0.6\\\hline
P9 & 31.94 & 0.38 & 16.97 & 7.66 & 11.35 & 0.66\\\hline
P10 & 32.38 & 0.84 & 13.81 & 1.91 & 7.1 & 4.6 \\ \hline
P11 & 15.65 & 0.34 & 2.93 & 2.21 & 3.79 & 0.6\\\hline
P12 & 10.5 & 0.19 & 13.49 & 0.5 & 3.71 & 0.27 \\ \hline \hline
\textbf{average} & 17.64 & 0.37 & 12.23 & 3.82 & 4.75 & 0.86\\
\bottomrule
\end{tabular}}
\label{individual_results}
\vspace{-2mm}
\end{table}

Table~\ref{results} presents a performance comparison of MealMeter against three baseline methods. The results indicate that MealMeter outperforms the baseline methods across all three macronutrient categories in MAE and correlation. 

For carbohydrate estimation, MealMeter achieves the lowest MAE $(13.2$ g$)$ and RMSRE $(0.37)$ and highest Pearson correlation $(0.44)$. TabPFN achieves similar MAE score, however, its low correlation score is a concern. 

In protein estimation, MealMeter shows a substantial improvement with an MAE of $9.7$ g, whereas Huo et al., Yang et al. and TabPFN report much higher errors $(12.8$ g, $10.5$ g and $9.87$, respectively$)$. Similarly, RMSRE for MealMeter $(4.5)$ is lower than two baselines $(5.2$ and $7.2)$, yet, remains high in capturing protein intake. The relatively high error in protein estimation suggests that wristband and CGM-based multimodal data may not be sufficiently informative for accurately capturing protein intake.

MealMeter also demonstrates better accuracy for fat estimation, with an MAE of $3.67$ g, significantly lower than Huo et al. $(4.8$ g$)$, Yang et al. $(4.7$ g$)$ and TabPFN $(3.9$ g$)$. The RMSRE for fat $(0.74)$ is lower than two baselines $(0.82$ and $1.02)$ but higher than that of TabPFN $(0.63)$, indicating reduced relative error.

The Pearson correlations are close to $0.45$ for all macronutrient types. Figure~\ref{r_plot} depicts scatter plot for predicted and actual carbohydrate amounts along with regression line and 95\% confidence interval. TabPFN suffers from low correlation as the dataset is too smalls to fine-tune it.

Therefore, the aforementioned results highlight MealMeter’s ability to enhance macronutrient estimation accuracy. Furthermore, MealMeter is faster than the baselines, thanks to its lightweight machine learning model.

\begin{table*}[!h]
\centering
\small
\caption{Comparing MealMeter against the baselines using MAE (grams), RMSRE and Pearson correlation}
{\renewcommand{\arraystretch}{1.3}
\begin{tabular}{|C{2.1cm}|C{1.1cm}C{1.45cm}C{0.9cm}|C{1.1cm}C{1.45cm}C{0.9cm}|C{1.1cm}C{1.45cm}C{0.9cm}|}
\toprule 
   & \multicolumn{3}{c|}{\textbf{Carbohydrate}} & \multicolumn{3}{c|}{\textbf{Protein}} & \multicolumn{3}{c|}{\textbf{Fat}} \\ \hline
  \textbf{Method} & \cellcolor{blue!18}\textbf{MAE $\downarrow$} & \cellcolor{gray!18}\textbf{RMSRE $\downarrow$} & \cellcolor{red!25}\textbf{\textit{r $\uparrow$}} & \cellcolor{blue!18}\textbf{MAE $\downarrow$} & \cellcolor{gray!18}\textbf{RMSRE $\downarrow$} & \cellcolor{red!25}\textbf{\textit{r $\uparrow$}} & \cellcolor{blue!18}\textbf{MAE $\downarrow$} & \cellcolor{gray!18}\textbf{RMSRE $\downarrow$} & \cellcolor{red!25}\textbf{\textit{r $\uparrow$}} \\ \hline
MealMeter & \textbf{13.2} & \textbf{0.37} & \textbf{0.44} & \textbf{9.66} & 4.51 & \textbf{0.43} & \textbf{3.67} & 0.74 & \textbf{0.49} \\ \hline
Huo et al. \cite{Huo2019PredictingTM} & 14.8 & 0.5 & 0.36 & 12.8 & 5.2 & 0.2 & 4.8 & 0.82 & 0.29 \\\hline
Yang et al.\cite{Yang2021AML} & 17 & 0.51 & 0.42 & 10.5 & 7.2 & 0.42 & 4.7 & 1.02 & 0.41 \\  \hline
TabPFN \cite{hollmann2023tabpfn} & \textbf{13.2} & 0.44 & 0.12 & 9.87 & \textbf{3.61} & 0.1 & 3.91 & \textbf{0.63} & 0.06 \\ \hline
\bottomrule
\end{tabular}}
\label{results}
\vspace{-2mm}
\end{table*}

The bar charts in Figure~\ref{contribution} illustrate the contributions of different physiological signals towards the estimation of macronutrient intake. 

EDA and postprandial BGL signals exhibited the highest contributions for carbohydrate estimation, followed by skin TEMP and HR. Acc showed a slightly negative contribution, which suggestis limited predictive value.

EDA remained the most influential signal for protein estimation, while contributions from HR, skin TEMP, and BGL signals were relatively balanced. Acceleration had a negligible impact once again.

Postprandial BGLs showed the highest contribution in Fat Estimation, with moderate influence from skin TEMP and EDA. Other signals had comparatively lower contributions.

\section{Conclusion}

MealMeter has achieved notable performance in meal tracking showcasing its potential for dietary assessment. However, our long-term vision extends beyond meal tracking in controlled environments. We aim to enhance MealMeter’s capabilities for more realistic applications. To implement this plan, we are actively collecting similar multimodal data in free-living settings, where participants systematically log their food intake while going about their daily routines, and we impose no restriction on their mobility. Furthermore, we plan to integrate a set of $\mu$EMA prompts to assess participants’ stress levels at 30-minute intervals. With this additional modality, our goal is to investigate the relationship between stress and dietary intake and enhance the model’s ability for dietary assessment and metabolic health management.

\addtolength{\textheight}{-12cm}   




\section*{ACKNOWLEDGMENT}

This work was supported by the National Institute of Diabetes and Digestive and Kidney Diseases of the National Institutes of Health under Award Number T32DK137525. Any opinions, findings, conclusions, or recommendations expressed in this material are those of the authors and do not necessarily reflect the views of the funding organization.

\bibliographystyle{IEEEtran}

\end{document}